\documentclass[final]{svjour2}
\usepackage{graphicx}
\usepackage{rotating}
\usepackage{amssymb}
\usepackage{mathptmx}
\usepackage{bm}
\usepackage[numbers]{natbib}

\def\lesssim{\ \raise.3ex\hbox{$<$}\kern-0.8em\lower.7ex\hbox{$\sim$}\ }
\def\gesim{\ \raise.3ex\hbox{$>$}\kern-0.8em\lower.7ex\hbox{$\sim$}\ }
\def\dirac{\ \raise.6ex\hbox{$-$}\kern-0.6em\hbox{$h$}\ }

\makeatletter
\journalname{Journal of Low Temperature Physics}
\bibpunct{}{}{,}{s}{}{,}

\begin{document}

\newcommand{\hdblarrow}{H\makebox[0.9ex][l]{$\downdownarrows$}-}
\title{Two-dimensional pseudogap effects of an ultracold Fermi gas in the BCS-BEC crossover region}

\author{R. Watanabe$^1$ \and S. Tsuchiya$^{2,3}$ \and Y. Ohashi $^1$}

\institute{1:Department of Physics, Faculty of Science and Technology, Keio University, Yokohama, Japan\\
\email{rwatanab@rk.phys.keio.ac.jp}
\\2: Department of Physics, Faculty of Science, Tokyo University of Science, Tokyo, Japan\\ 3: Research and Education Center for Natural Sciences, Keio University, Yokohama, Japan}

\date{07.07.2012}

\maketitle

\keywords{cold Fermi gas, pseudogap, low-dimensionality}

\begin{abstract}
We investigate pseudogap phenomena originating from pairing fluctuations in the BCS-BEC crossover regime of a two-dimensional Fermi gas in a harmonic trap. Including pairing fluctuations within a $T$-matrix theory and effects of a trap within the local density approximation, we calculate the local density of states (LDOS) at the superfluid phase transition temperature $T_{\rm c}$. In the weak-coupling regime, we show that the pseudogap already appears in LDOS around the trap center. The spatial region where the pseudogap can be seen in LDOS becomes wider for a strong pairing interaction. We also discuss how the pseudogap affects the spectrum of the photoemission-type experiment developed by JILA group.

PACS numbers: 03.75.Hh, 05.30.Fk, 67.85.Lm.
\end{abstract}

\section{Introduction}
Ultracold Fermi atom gases are now widely recognized as a powerful quantum system to study many-body physics in strongly interacting fermion systems\cite{GIORGINI,BLOCH}. One typical example is the BCS (Bardeen-Cooper-Schrieffer)-BEC (Bose-Einstein condensation) crossover phenomenon realized in $^{40}$K and $^6$Li Fermi gases, where the character of superfluidity continuously changes from the BCS-type to the BEC-type, with increasing the strength of a pairing interaction associated with a Feshbach resonance\cite{GIORGINI,BLOCH,OHASHI1}. Using this phenomenon, we can study Fermi superfluids and Bose superfluids in a unified manner. The intermediated coupling regime, which is sometimes referred to as the crossover region, has attracted much attention, because various anomalous phenomena originating from strong pairing fluctuations are expected there. Indeed, anomalous single-particle excitation spectra have been recently observed in a $^{40}$K Fermi gas by using a photoemission-type experiment\cite{STEWART,GAEBLER}. Theoretically, it has been pointed out\cite{TSUCHIYA,WATANABE,TSUCHIYA2,TSUCHIYA3} that this anomaly can be quantitatively understood as a pseudogap phenomenon associated with strong pairing fluctuations. 
\par
Although the origin of the observed anomalous single-particle excitation spectra is still in debate, the above mentioned pseudogap scenario is a strong candidate. In this regard, to assess the validity of this idea, a two-dimensional Fermi gas would be useful, because stronger pairing fluctuations than the three-dimensional case are expected. Recently, the photoemission-type experiment has been done for a quasi-two-dimensional $^{40}$K Fermi gas\cite{FELD}. In this experiment, in addition to the so-called back-bending curve, the split of the spectrum is also observed in the crossover region, which is quite different from the three-dimensional result (where a single broad peak is only observed).
\par
In this paper, we investigate single-particle excitations in the BCS-BEC crossover regime of a two-dimensional Fermi gas. Although the recent experiment deals with a {\it quasi}-two dimensional gas\cite{FELD}, we ignore the $z$-direction, for simplicity. To examine the validity of the pseudogap scenario proposed in the three-dimensional case, we include strong-coupling pairing fluctuations within a $T$-matrix theory in the normal state. We also include effects of a harmonic trap within the local density approximation (LDA). Using this combined $T$-matrix theory with LDA, we calculate the local density of states (LDOS), as well as the photoemission spectra, at the superfluid phase transition temperature $T_{\rm c}$. We note that this framework has succeeded in quantitatively explaining the observed photoemission spectra in a three-dimensional $^{40}$K Fermi gas\cite{TSUCHIYA2,TSUCHIYA3}. We also note that, in the presence of a trap, the present $T$-matrix theory with LDA gives a finite $T_{\rm c}$, as in the BEC case of a trapped Bose gas\cite{Pethick}. In the absence of trap, the superfluid phase transition in a {\it uniform} two-dimensional Fermi gas is dominated by the Kosterlitz-Thouless transition\cite{KT}.
\par
Throughout this paper, we set $\dirac=k_{\rm B}=1$, and the system volume $V$ is taken to be unity.
\par
\section{Two-dimensional trapped Fermi gas in the BCS-BEC crossover}
\par
We consider a two-dimensional Fermi gas, described by the BCS Hamiltonian,
\begin{equation}
H=\sum_{{\bm p},\sigma}\xi_{\bm p}c_{{\bm p}\sigma}^\dagger c_{{\bm p}\sigma}-U\sum_{\bm {p,p^\prime,q}}c_{{\bm {p+q/2}}\uparrow}^\dagger c_{{\bm {-p+q/2}}\downarrow}^\dagger c_{{\bm {-p^\prime+q/2}}\downarrow}c_{{\bm {p^\prime+q/2}}\uparrow}.
\end{equation} 
Here, $c_{{\bm p}\uparrow}^\dagger$ is a creation operator of a Fermi atom with pseudo-spin $\sigma=\uparrow,\downarrow$, describing two atomic hyperfine states. $\xi_{\bm p}=\epsilon_{\bm p}-\mu=\frac{p^2}{2m}-\mu$ is the kinetic energy measured from the Fermi chemical potential $\mu$, where $m$ is an atomic mass. The pairing interaction $-U (<0)$ is assumed to be tunable. As usual, we measure the interaction strength in terms of the $s$-wave scattering length $a_s$. In the two-dimensional case, the scattering length is related to $-U$ as\cite{PIETILA,MORGAN}
\begin{equation}
-U^{-1}=\frac{m}{2\pi}\ln\frac{2}{C\sqrt{-2mE_+}a_s}-\sum_{\bm p}\frac{1}{E_++i\delta-2\epsilon_p},
\label{eq.A}
\end{equation}
where $\delta=+0$, and $C=e^{\gamma}$ (where $\gamma=0.577$ is the Euler's constant). In Eq. (\ref{eq.A}), $E_+$ is a small positive energy, which is taken to be zero in the final expressions for the $T$-matrix theory. 
\par
\begin{figure}
\begin{center}
\includegraphics[%
  width=.9\linewidth,
  keepaspectratio]{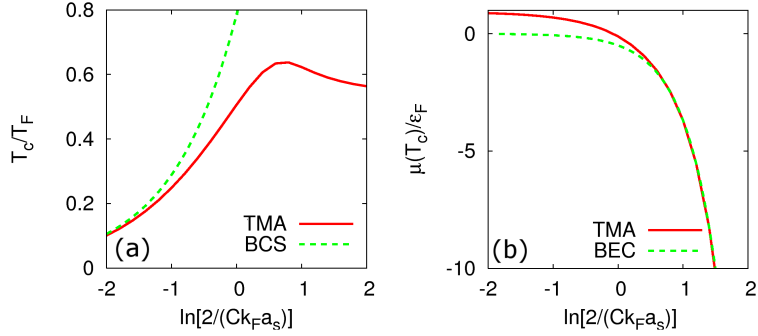}
\end{center}
\caption{(color online)Calculated $T_{\rm c}$ (a) and $\mu(T=T_{\rm c})$ (b) in the BCS-BEC crossover regime of a trapped two-dimensional Fermi gas. The scattering length $a_s$ is normalized by the Fermi momentum $k_{\rm F}$. $T_{\rm c}$ and $\mu(T=T_{\rm c})$ are normalized by the Fermi temperature $T_{\rm F}$ and the Fermi energy $\varepsilon_{\rm F}$, respectively. `TMA' and `BCS' show the results based on the $T$-matrix theory with LDA and the mean-field theory, respectively. In panel (b), `BEC' shows the result based on the theory valid for the strong-coupling BEC limit.
}
\label{FIG1}
\end{figure}

Figure \ref{FIG1} shows the calculated $T_{\rm c}$ and $\mu(T=T_{\rm c})$. In panel (a), $T_{\rm c}$ continuously changes from the weak-coupling BCS result $\frac{2}{\pi\sqrt{m}a_s}T_{\rm F}$ to that for a molecular Bose gas\cite{BLOCH} $\frac{\sqrt{3}}{\pi}T_{\rm F}$ (where $T_{\rm F}$ is the Fermi temperature). In this BCS-BEC crossover, the chemical potential $\mu$ at $T_{\rm c}$ changes from $\mu=\epsilon_{\rm F}$ to $-\frac{1}{C^2ma_s^2}$, as shown in panel (b). Figure \ref{FIG1} indicates that the character of superfluidity changes from the BCS-type to the BEC of bound molecules around $-1\lesssim\ln\frac{2}{Ck_{\rm F}a_s}\lesssim+1$. Thus, although there is no clear boundary between the weak-coupling BCS regime and the strong-coupling BEC regime, in what follows, we call the region $\ln\frac{2}{Ck_{\rm F}a_s}<0$ and $\ln\frac{2}{Ck_{\rm F}a_s}>0$ the BCS side and BEC side, respectively.
\par
Effects of pairing fluctuations are conveniently described the self-energy correction $\Sigma_{\bm p}(i\omega_n)$ in the single-particle Green's function,
\begin{equation}
G_{\bm p}(i\omega_n)=
{1 \over G_{\bm p}^0(i\omega_n)^{-1}-\Sigma_{\bm p}(i\omega_n)}, 
\label{eq.B}
\end{equation}
where $\omega_n$ is the fermion Matsubara frequency, and $G_{\bm p}^0(i\omega_n)=(i\omega_n-\xi_{\bm p})^{-1}$ is the Green's function for a free Fermi gas. In the $T$-matrix approximation, the self-energy $\Sigma_{\bm p}(i\omega_n)$ is diagrammatically given by the sum of ladder-type diagrams. The resulting expression is\cite{TSUCHIYA,PIETILA}
\begin{equation}
\label{SIGMA}
\Sigma_{\bm p}(i\omega_n)=T\sum_{{\bm p},\nu_n}\Gamma_{\bm q}(i\nu_n)G_{\bm {q-p}}^0(i\nu_n-i\omega_n).
\end{equation}
Here, $\nu_n$ is the boson Matsubara frequency. The particle-particle scattering matrix $\Gamma_{\bm q}(i\nu_n)$ describes pairing fluctuations in the Cooper channel, given by
\begin{equation}
\Gamma_{\bm q}(i\nu_n)=-{U \over 1+U\Pi_{\bm q}(i\nu_n)},
\label{eq.C}
\end{equation}
where $\Pi_{\bm q}(i\nu_n)=T\sum_{\bm{p},\omega_n}G_{\bm {p+q/2}}^0(i\nu_n+i\omega_n)G_{\bm {-p+q/2}}^0(i\omega_n)$ is the lowest order pair-propagator.
\par
We now include effects of a trap potential. For simplicity, we assume an isotropic harmonic trap (with the trap frequency $\Omega$) as $V(r)=m\Omega^2r^2/2$. Within LDA, effects of this trap can be conveniently incorporated into the theory by simply replacing the Fermi chemical potential $\mu$ with $\mu(r)=\mu-V(r)$. Then, the LDA Green's function $G_{\bf p}(i\omega_n)$, self-energy $\Sigma_{\bf p}(i\omega_n)$, and the scattering matrix $\Gamma_{\bf q}(i\nu_n)$, depend on $r$ through the position-dependent LDA chemical potential $\mu(r)$. The chemical potential $\mu$ is determined by the LDA equation for the number $N$ of Fermi atoms, given by
\begin{equation}
N=2\pi \int rdr n(r),
\label{eq.D}
\end{equation}
where $n(r)=T\sum_{{\bm p}\sigma}G_{{\bm p}\sigma}(i\omega_n)$ is the number density of Fermi atoms in LDA.
\par
As in the three-dimensional case, we determine $T_{\rm c}$ from the Thouless criterion, which states that the superfluid phase transition occurs when the LDA scattering matrix has a pole at ${\bf q}=\nu_n=0$. In LDA, this condition is first satisfied in the trap center, so that one may solve $\Gamma_{{\bf q}=0}(i\nu_n=0,r=0)^{-1}=0$ to determine $T_{\rm c}$. We actually solve this $T_{\rm c}$-equation, together with the number equation (\ref{eq.D}). We briefly note that, because of the presence of a trapped potential, although the system itself is a two-dimensional one, a finite $T_{\rm c}$ is obtained within the present combined $T$-matrix approximation with LDA\cite{Pethick}. 

\par

\section{Local density of states and photoemission spectrum}
\par
Once Fig.\ref{FIG1} is obtained, the LDA local density of states $\rho(\omega,r)$ at $T_{\rm c}$ can be calculated from the LDA Green's function $G_{\bf p}(i\omega,\mu\to\mu(r))$ as
\begin{equation}
\rho(\omega,r)=\sum_{\bm p}A_{\bm p}(\omega,r),
\label{eq.E}
\end{equation}
where $A_{\bm p}(\omega,r)=-\pi^{-1}{\rm Im}[G_{{\bm p}\sigma}(i\omega_n\to\omega+i\delta,r)]$ is the single-particle spectral weight. The photoemission spectrum $\overline{A_{\bm p}(\omega)f(\omega)}$ is also related to the spectral weight $A_{\bm p}(\omega,r)$ as\cite{TSUCHIYA3}
\begin{equation}
\overline{A_{\bm p}(\omega)f(\omega)}=\frac{2 t_{\rm F}^2}{R_{\rm F}^2}\int d{\bf r}A_{\bm p}(\omega+V(r),r)f(\omega+V(r)),
\label{eq.F}
\end{equation}
where $f(\omega)$ is the Fermi distribution function, and $R_{\rm F}$ is the LDA radius of the gas cloud at $T=0$ for a free Fermi gas with $N$ atoms. $t_{\rm F}$ is a rf-coupling constant. Since the current photoemission experiments do not have enough spatial resolution, the spatial average is taken in Eq. (\ref{eq.F}).
\par
\begin{figure}
\begin{center}
\includegraphics[%
  width=.8\linewidth,
  keepaspectratio]{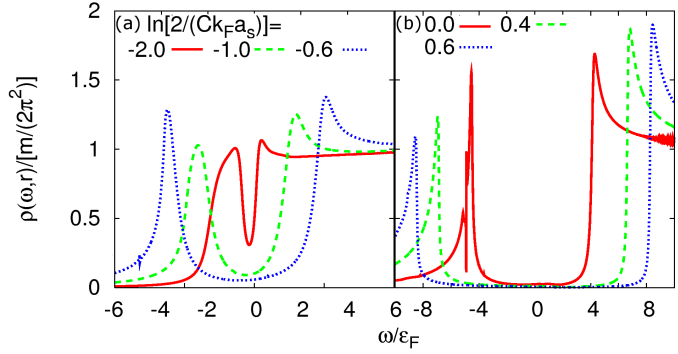}
\end{center}
\caption{(color online)Calculated local density of states $\rho(\omega,r)$ at $r=0.01R_{\rm F}$ when $T=T_{\rm c}$. (a) BCS side ($\ln\frac{2}{Ck_{\rm F}a_s}\le 0$). (b) BEC side ($\ln\frac{2}{Ck_{\rm F}a_s}\ge 0$).}
\label{FIG2}
\end{figure}

Figure \ref{FIG2} shows the local density of states $\rho(\omega,r)$ near the trap center ($r=0.01R_{\rm F}$) when $T=T_{\rm c}$\cite{note1}. In the weak-coupling BCS regime shown in panel (a), one already sees a dip structure around $\omega=0$. Since the superfluid order parameter vanishes at $T_{\rm c}$, this is just the pseudogap associated with pairing fluctuations. Since $T_{\rm c}$ is close to the mean-field value when $\ln\frac{2}{Ck_{\rm F}a_s}=-2.0$ (See Fig.\ref{FIG1}.), this result is quite different from the three-dimensional case, where the pseudogap does not appear in such a weak-coupling regime. This is simply because of a stronger pairing fluctuations existing in the two-dimensional system.
\par
This pseudogap structure becomes remarkable, as one increases the interaction strength. In the BEC side shown in Fig.\ref{FIG2}(b), the overall structure is very close to the BCS superfluid density of states, characterized by a finite energy gap with two coherence peaks at the excitation edges. 
\par
\begin{figure}
\begin{center}
\includegraphics[%
  width=\linewidth,
  keepaspectratio]{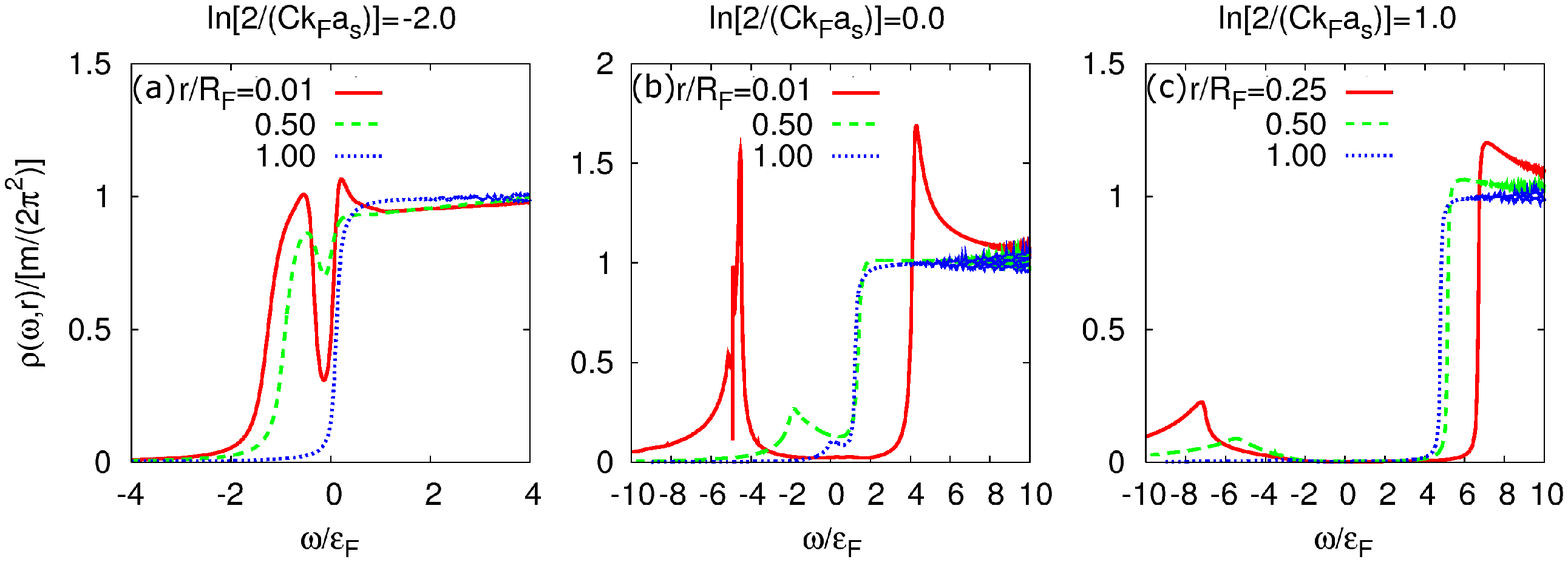}
\end{center}
\caption{(color online)Spatial variation of the local density of states $\rho(\omega,r)$ at $T_{\rm c}$. The interaction strength $\ln\frac{2}{Ck_{\rm F}a_s}$ is set to be (a) -2, (b) 0, and (c) +1. $R_{\rm F}$ is the LDA radius of a free gas of $N$ Fermi atoms at $T=0$.}
\label{FIG3}
\end{figure}
\par
Figure \ref{FIG3} shows the $r$-dependence of LDOS at $T_{\rm c}$. In the weak-coupling BCS regime shown in panel (a), the pseudogap structure seen in the trap center gradually disappears in the outer region of the gas cloud. In addition, the peaks around the edges of the pseudogap is suppressed when $r/R_{\rm F}\ge 0.5$. While a similar spatial dependence of LDOS is still obtained in the crossover region (Fig.\ref{FIG3}(b)), Fig.\ref{FIG3}(c) shows that a large gap structure remains to $r=R_{\rm F}$ in the BEC regime. In this strong-coupling regime, since the Fermi chemical potential $\mu$ is negatively large (See Fig.\ref{FIG1}(b).), the system is expected to be well described by a gas of two-body bound molecules. Thus, the large gap structure seen in Fig.\ref{FIG3}(c) is considered to originate from a molecular binding energy. 
\par

\begin{figure}
\begin{center}
\includegraphics[%
  width=\linewidth,
  keepaspectratio]{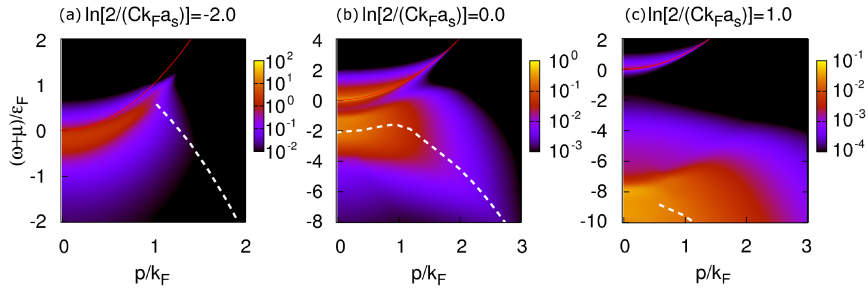}
\end{center}
\caption{(color online)Intensity of photoemission spectrum $\overline{A_{\bm p}(\omega)f(\omega)}$ at $T_{\rm c}$. (a) BCS regime ($\ln\frac{2}{Ck_{\rm F}a_s}=-2$). (b) crossover regime ($\ln\frac{2}{Ck_{\rm F}a_s}=0$). (c) BEC regime ($\ln\frac{2}{Ck_{\rm F}a_s}=1$). In each panel, the dashed line shows the position of the back-bending curve in the spectrum. The solid line shows the free-particle dispersion, $\omega+\mu=\epsilon_{\bm p}$.
}
\label{FIG4}
\end{figure}

Figure \ref{FIG4} shows the photoemission spectra in the BCS-BEC crossover at $T_{\rm c}$. In the weak-coupling BCS regime (panel (a)), the peak position in the spectral intensity is close to the free particle dispersion $\omega+\mu=p^2/(2m)$. In the crossover region (panel (b)), in addition to this free particle dispersion, one obtains another branch in the negative energy region. In the strong coupling BEC regime (panel (c)), a large gap structure is seen in the photoemission spectrum, reflecting a large binding energy of a bound molecule. 
\par
We note that, in Fig.\ref{FIG4}(b), while the upper free-particle-like branch mainly comes from single-particle excitations around the edge of the gas (where the pseudogap effect is almost absent), the lower branch is dominated by the pseudo-gapped local density of states around the trap center. This lower branch exhibits the so-called back-bending behavior (dashed line in Fig.\ref{FIG4}(b)), which is consistent with the recent experiments on a quasi-two-dimensional $^{40}$K Fermi gas\cite{FELD}. We briefly note that such a back-bending curve can be already slightly seen in the weak-coupling BCS regime (dashed line in Fig.\ref{FIG4}(a)).
\par
\section{Summary}
To summarize, we have discussed pseudogap phenomena originating from strong pairing fluctuations in a trapped two-dimensional Fermi gas. Using a combined $T$-matrix theory with LDA, we have examined how the pseudogap appears in the local density of states, as well as the photoemission spectrum, in the entire BCS-BEC crossover region at $T_{\rm c}$. We showed that the photoemission spectrum splits into the upper and lower branches in the crossover region. The lower branch, which is deeply related to the pseudo-gapped local density of states around the trap center, exhibits a back-bending behavior, which is consistent with the recent photoemission experiment on a quasi-two-dimensional Fermi gas. Since two-dimensional Fermi gases have recently attracted much attention, our results would be useful for the further development of ultracold Fermi gas physics.

\begin{acknowledgements}
We would like to thank S. Watabe, T. Kashimura, D. Inotani, and R. Hanai for fruitful discussions. R.W. was supported by the Japan Society for the Promotion of Science. Y. O. was supported by Grant-in-Aid for Scientific research from MEXT in Japan (22540412, 23104723, 23500056).
\end{acknowledgements}


\begin{thebibliography}{99}
\bibitem{note1} At $T_{\rm c}$, the LDA scattering matrix diverges as $\Gamma_{\bm q}(i\nu_n=0,r=0)\sim q^{-2}$ ($q\to0$), because the Thouless criterion is satisfied at this temperature. In the two-dimensional case, this leads to the diverging behavior of the analytic-continued LDA self-energy at $r=0$, so that one cannot calculate $\rho(\omega,r=0)$ at $T_{\rm c}$. We note that, although the number density $n(r)$ also diverges in the trap center because of the same origin, the total number $N$ of Fermi atoms in Eq. (\ref{eq.D}) converges.
\end{thebibliography}
\end{document}